\documentclass[11pt]{article}
\renewcommand{\baselinestretch}{1.5}

\usepackage{amsmath}
\usepackage{color}
\usepackage{latexsym}
\usepackage{graphicx,graphics,subfigure}
\DeclareGraphicsExtensions{.bmp,.jpg,.eps,.pdf}
\usepackage{float,verbatim}

\begin{document}
\title{Sequential Design for Computer Experiments with a Flexible Bayesian
Additive Model}
\author{Hugh Chipman, Pritam Ranjan and Weiwei Wang \vspace{0.2cm}\\
\small Department of Mathematics and Statistics,
Acadia University, NS, Canada\\ [-0.2cm]
\small (hugh.chipman@acadiau.ca, pritam.ranjan@acadiau.ca, 084684w@acadiau.ca)
}
\date{}
\maketitle
\begin{abstract}
In computer experiments, a mathematical model implemented on a computer
is used to represent complex physical phenomena.  These models, known as computer
simulators, enable experimental study of a virtual representation of
the complex phenomena.  Simulators can be thought of as complex functions
that take many inputs and provide an output.  Often these simulators are
themselves expensive to compute, and may be approximated by ``surrogate
models'' such as statistical regression models.  In this paper we consider
a new kind of surrogate model, a Bayesian ensemble of trees (Chipman et al.
2010), with the
specific goal of learning enough about the simulator that a particular
feature of the simulator can be estimated.  We focus on identifying
the simulator's global minimum.  Utilizing the Bayesian version of the
Expected Improvement criterion (Jones et al.  1998), we show that this
ensemble is particularly effective when the simulator is ill-behaved,
exhibiting nonstationarity or abrupt changes in the response.  A number
of illustrations of the approach are given, including a tidal power
application.\\


\end{abstract}
\renewcommand{\baselinestretch}{1.4}

\section{Introduction}\label{sec:intro}

Many phenomena, such as tidal flow, nuclear reactions, climate behaviour and universe expansion are sufficiently large and complex that direct scientific experiments are
either impossible or impractical, due to time and cost constraints.  As a
result, mathematical models are often used to build a realistic
representation of these phenomena, enabling experimentation.  For example,
with a model of tidal flow, researchers can study the effect of placing
underwater electrical turbines, using the mathematical model to
``observe'' how tidal flow would change, and predict how much electrical
power might be generated.

These mathematical models, often known as computer simulators, typically
take a number of inputs and when they are run, generate an output.  Running
computer simulators
can be time-consuming when output values
are desired for a large number of different input values.
In such situations, a second level of
approximation, in which a ``surrogate model'' is used to approximate the
input-output relationship of the simulator, is often used.  These surrogate
models are flexible regression models, taking an input vector $x$ and
predicting an output $y$.

As with most statistical models, a surrogate model is estimated with
training data, consisting of $n$ observations $(x_1,y_1), \ldots,
(x_n,y_n)$.  Unlike many real-world phenomena, a computer simulator is
often deterministic.  That is, every time the simulator is run with input
$x$, the same numeric value of output $y$ will result.  Surrogate models
that capture such behaviour, interpolating exactly the training data,
are popular tools in such computer experiments.

In this paper we consider computer experiments with the very specific goal
of optimization.  We wish to find the values of input $x$ which minimize (or
maximize) the output $y$.  This sort of approach is taken in
Jones, Schonlau and Welch (1998), where the training data are collected
sequentially, allowing refinement of the surrogate model and increasingly
accurate prediction of the location of the global minimizer.

By far the most popular statistical surrogate model is the Gaussian Process
(GP) model (for an overview, see Santner, Williams and Notz
(2003) and Rasmussen and Williams (2006)).  The GP assumes a multivariate normal distribution for a vector of
responses at corresponding input locations.  The covariance of the
responses is taken to depend on the distances between the input locations.
Closely related to spatial models (e.g., kriging), the GP model allows
interpolation of a deterministic function, and can predict quite
effectively in the input space.

The GP model makes rather strong assumptions, however.  The assumption that
covariance depends only on distance between points and not their location  leads to a stationary model, in which the
smoothness of the response function is the same over the entire input
space.  Limitations such as this have led to extensions.  For example,
Gramacy and Lee (2008) develop a treed Gaussian Process (TGP) model, in
which a tree recursively partitions the input space into rectangular regions, and
within each region, a stationary GP model is fit.

In this paper we consider a more radical departure from the GP model.
We build upon a Bayesian {\em ensemble of trees} model (Chipman, George
and McCulloch 2010, here onwards referred to as CGM), which combines the outputs from a large number of regression trees via a summation.  This gives a flexible model capable of
capturing nonstationarity and complex relationships, and if present,
additive structure.  It is also able to emulate both deterministic
and stochastic (noisy) simulators. By utilizing Markov chain Monte Carlo
(MCMC) for
estimation, full inference for the model is available.  This in turn
enables computation of merit based criteria, for instance, an Expected Improvement (EI) criterion (Jones et al. 1998), that can guide the sequential design of experiments for
identification of a global minimum (our focus in this article) or other
features of interest.

The remainder of the paper is organized as follows.  In Section 2,
we introduce Bayesian Additive Regression Trees (BART), the ensemble
that will be used as the surrogate model.  After discussing adaptation
of prior distributions for BART to computer experiments, we briefly
outline how TGP may be used as another surrogate model.  The two models
are illustrated with a tidal power generation example.  In Section 3,
the Expected Improvement criterion of Jones et al. (1998) is reviewed
and adapted to the BART model.  A number of illustrations are given
in Section 4, with test functions playing the role of computer models,
and up to four-dimensional inputs considered.  Section 5 concludes with
a discussion of future directions for research.

\section{Statistical Surrogate Models} \label{sec:models}
In this section, we present a brief review of the BART model (see CGM
for more details).  We also discuss the Treed Gaussian Process (TGP)
model of Gramacy \& Lee (2008), which will be compared to BART in
Section~\ref{sec:results}.

\subsection{The BART model} \label{sec:bartmodel}
Assume the simulator takes a $d$-dimensional input vector $x=(x_1, \ldots, x_d)$
and has real-valued output $y(x)$. The BART model represents the output
as a sum of $m$ adaptively chosen functions and an independent normal error,
\begin{equation}\label{eq:bart}
y(x) = \sum_{j=1}^m g(x; T_j,M_j)\,  + \,
\epsilon = h(x) + \epsilon ,
 \qquad \epsilon \sim N(0,\sigma^2).
\end{equation}
The function $g(x;T,M)$ produces an output when provided with input vector $x$
and parameters $T$ and $M$.  It is the shorthand notation for a ``regression
tree'' model. The predictions for a particular value of $x$ are generated by following the
sequence of decision rules in tree $T$ until arriving at a terminal node $b$,
at which point an associated scalar prediction $\mu_b$ is returned.
Decision rules in $T$ branch on a single element of the vector $x$, say
$x_1$, yeilding rules such as $x_1 < 2.4$.  Different rules in each tree
may use different variables.  For a
tree $T$ with $B$ terminal nodes (i.e., partitioning the input space into
$B$ rectangular regions), let $M=(\mu_1,\ldots, \mu_B)$ denote the
collection of terminal node predictions.

Thus, viewed as a function of $x$, tree model $g$ produces a piecewise-constant
output.  By combining together an ``ensemble'' of $m$ such tree models in
(\ref{eq:bart}), a flexible modelling framework is created.  For
instance, if each individual $T_j$ uses partitions on only a single
input variable, then the BART model becomes an additive model.  BART is
however capable of incorporating higher-dimensional interactions, by
adaptively choosing the structure and individual rules of the $T_j$'s.  Furthermore, many
individual trees ($T_j$) may place split points in the same area, allowing
the predicted function to change rapidly nearby, effectively capturing
nonstationary behaviour such as abrupt changes in the response.

Viewing this as a statistical model, we have parameters $\Theta
=(T_1,\ldots T_m, M_1, \ldots, M_m, \sigma)$.  CGM take quite large values of $m$ (say,
50 to 200), and estimate (\ref{eq:bart}) in a Bayesian framework using
MCMC.  With a sufficiently large number of terms
in the ensemble, the interpretation of individual trees becomes irrelevant,
and the most useful interpretation of BART is to
view it as a way of placing a prior on $E(Y|X)$, that is a prior on
functions.  Of course, all regression models, including GPs, place a prior
on functions.  What is interesting is the sort of functions that receive
prior mass.

The BART model places a
prior on functions that differs from GPs in several important ways.  First,
as mentioned above, the BART model is biased towards additive and
low-dimensional functions (i.e. $g(x;T_j,M_j)$ in (\ref{eq:bart}) with small
trees $T_j$, that are
functions of one or a few variables). Second,
the BART model does not assume
continuity of the response, thus making it appropriate when there are
abrupt changes or non-stationarity in the response.
In contrast, the explicit specification of a spatial correlation structure
in a GP model implies continuity and, for many common
correlation functions, stationarity (i.e., a constant
amount of ``wiggle'') of the response.

The Bayesian framework equips the BART model with a full suite of
inferential tools. Most importantly, uncertainty in predicting $h(x)$
due to all parameters is available.  As will be demonstrated in Section 3,
this is a key ingredient in sequential design.

\subsection{Prior specification of BART for computer experiments}
\label{sec:bartprior}
CGM demonstrate effective performance of BART in a wide range of simulated and
real-data examples. In all examples, there was considerable noise,
that is $\sigma \gg 0$.  Often in computer experiments, the simulator is
deterministic ($\sigma =0$) or has small errors ($\sigma \approx 0$).  This
property underlies some important changes to CGM's default choice
of prior distributions.  This section briefly outlines
those choices, with emphasis on modifications for surrogate modelling of the deterministic simulator output.

We follow the same general formulation of prior distributions as in CGM,
with different parameter choices. CGM simplify prior specification
by assuming that, a priori, i) $T_1, \ldots, T_m$ are i.i.d., ii) all
elements of $M_1, \ldots, M_m$ (i.e., the $\mu_{i,b}$, node $b$ of tree
$i$) are i.i.d. given all $T$'s, and $\sigma$ is independent of all $T$'s
and $M$'s.

A special prior for a generic output $\mu$ of a tree is used.
In a computer experiment, with $\sigma \approx 0$, we
believe $h(x) \approx y(x)$, that is, the overall output is quite
close to the observed response. Thus we choose a larger prior variance
for individual tree output $\mu_{b,i}$ than the CGM default.  This
allows the outputs more flexibility, giving $h(x) \approx y(x)$ at the
training points.
This is accomplished as follows: After scaling data values of $Y$
to the interval (-0.5, 0.5), CGM recommend a default prior $\mu\sim
N(0,\sigma^2_{\mu}) = N(0,1/(4k^2m))$ with $k=2$.
Since the prediction $h(x)$ in (\ref{eq:bart}) is a sum of $m$ distinct
$\mu$'s, the default prior means that $h(x) \sim N(0,1/(4k^2))$.
There is thus a prior probability of approximately 95\% that $h(x)$ falls
within $0 \pm 2\sigma_{\mu} = 0 \pm 1/k$, or (-0.5, 0.5) for $k=2$.  We
recommend relaxing this prior to $k=1$.
Of note is the fact CGM's default prior specification applies considerable
shrinkage to individual output $\mu_{i,b}$.  Choosing a smaller value of
$k$ (e.g., $k=1$) increases the prior variance of output $h(x)$, applying less
shrinkage (or smoothness) of the response.

The assumption that $\sigma \approx 0$ requires modification to the prior
on $\sigma$.  Essentially, we place most of the prior mass near 0.
This is accomplished with the same inverted-chi-squared prior for
$\sigma^2$ as in CGM, using their recommended value of 3 degrees of freedom, and anchoring the 90th percentile
of the $\sigma$ prior at {\tt 0.20$\times$sd(y)}, where {\tt sd(y)} is the
sample standard deviation of the training $y$ values.  The only
difference from the CGM default is the value used to anchor the 90th
percentile (i.e., {\tt 0.20$\times$sd(y)}).  This prior specification allows for some noise in the response
values.  This strategy facilitates MCMC mixing for BART, and can also be
considered as having a similar motivation to Gramacy and Lee (2012) who
advocate the inclusion of a residual error when approximating deterministic
functions with GPs, for numeric stability and predictive accuracy.

Although not part of the prior specification, several other operating
parameters of BART are chosen as follows:  The number of trees in the
ensemble is chosen as $m=100$.  Individual trees are allowed to split on a
fine grid of 1000 cutpoints along each axis. The MCMC algorithm uses 6000
iterations, discarding the first 2000 (burn in) and keeping every 20th thereafter, for
a sample of 200 posterior draws.  Larger posteriors
samples might be desirable, but with the quick mixing behaviour of BART
observed by CGM, a sample of this size will be sufficient for the
sequential design step covered in Section \ref{sec:bart-ei}.

The remaining elements of the prior specification involve the tree $T$.
We use defaults from CGM, with a prior on tree $T$ that puts probability mass of 0.05,
0.55, 0.28, 0.09 and 0.03 on trees with $1, 2, 3, 4$, and $\geq 5$ terminal
nodes.  While favouring small trees, this prior does not rule out the
possibility of very large trees.

\subsection{Treed Gaussian processes and Gaussian process models}\label{sec:tgp}
Later in the paper we shall make comparisons between BART and another
tree-based model, the treed Gaussian Process (TGP)
model of Gramacy and Lee (2008).  We briefly review the model and prior specification here.  The TGP model has form similar to (\ref{eq:bart}), but with
$m=1$, i.e. a single tree.  The replacement of the ensemble by a single
tree is offset by
an enrichment of the terminal node model.  Instead of a constant mean in
each terminal node, Gramacy and Lee assume that conditional on the tree
structure, the response $y(x)$ follows a GP model with var$(y(x)) = \sigma^2_h + \sigma^2$ and cov$(y(x), y(x')) = \sigma^2_hK(x-x')$, where $K(x-x')$ is the spatial correlation structure, and $\sigma^2_h$ and $\sigma^2$ are the process and noise variance respectively.  By allowing a distinct GP model for each terminal node, TGP can
accommodate nonstationarity in the response.

Since the TGP model was
developed for emulation of computer simulators, less modification is needed
of the prior settings. In all our experiments we used default parameters in
the R implementation (the {\tt tgp} package),
except for \verb+BTE = c(2000,6000,20), nug.p = c(1,10,1,10^5)+. The
BTE parameter indicates that the predictive samples are saved every
E $(20)$ MCMC rounds starting at round B $(2000)$, and stopping at T
$(6000)$. That is, TGP predictions were also based on 200 posterior draws,
the same as for BART. The {\tt nug.p}
parameter specifies a mixture of exponential priors (with expected values
$10^{-1}$ and $10^{-5}$) on the ``nugget'' ($\omega$), where $\sigma^2 =
\sigma^2_h\cdot \omega$.

Gaussian process (GP) models are also popular for sequential optimization.
Our comparisons also include GPs, using the implementation from the {\tt
tgp} package in R.  All settings for the GP model are as above for TGP,
except of course without the trees.

\subsection{Small illustration: 1D tidal power}\label{sec:tidal_power}
To give a taste of these two models, we briefly describe a tidal power
modelling application. A full description of the application is given
in
 Wang (2010) 
 and Ranjan et. al. (2011). The basic problem is to predict the amount of
kinetic energy (which in turn will be used for producing electricity) that
could be extracted from tidal flow through the Minas Passage in Nova
Scotia, Canada. This 7-km wide passage has some of the highest tides in the
world. According to Karsten et al. (2008), an individual turbine can
generate up to 1 MW of power, and  approximately 2.5 GW of power can be
harnessed from the tidal kinetic energy by placing large collections of turbines in the Minas Passage. Karsten et al. suggest that a ``fence'' of turbines across the passage is optimal for extracting the maximum possible energy at the minimum cost.

Wang 
considered the emulation of a simplified tidal flow model (simulator), in which fences of turbines are placed in the Minas Passage. The only parameter to control in this simulator is the location of the fence along the passage. That is, our ``$x$'', when
scaled to the interval $(0,1)$, represents the fence position along the
passage. Variation in the shape of the shoreline leads to considerable difference in the flow at different locations. A second very significant influence on flow is the placement
of other turbine fences. In the area surrounding a previously placed
turbine fence, the flow can be expected to be much lower.

\begin{figure}[h!]\centering
\subfigure[Realization~1]{\label{fig:BART-TGP-comparison:a} 
\includegraphics[height=3.6in,width=2.95in]{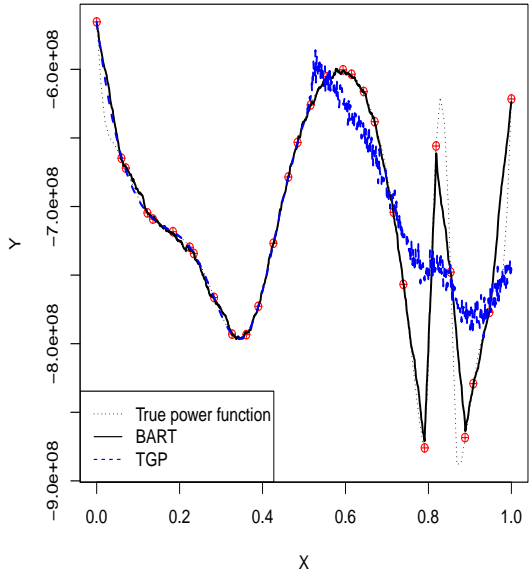}}
\subfigure[Realization~2]{\label{fig:BART-EI-comparison:b} 
\includegraphics[height=3.6in,width=2.95in]{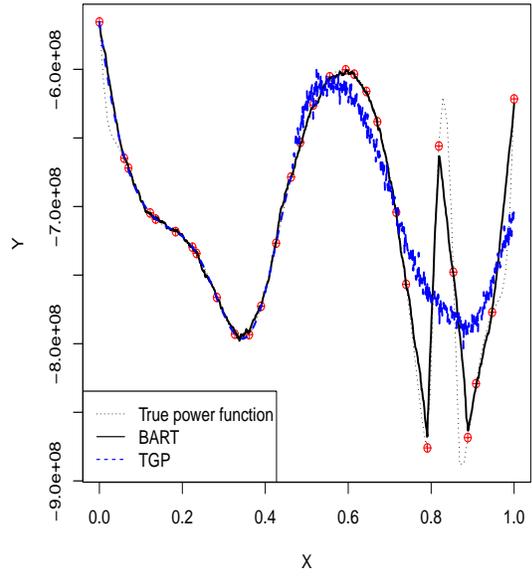}}
\subfigure[Realization~3]{\label{fig:BART-EI-comparison:c} 
\includegraphics[height=3.6in,width=2.95in]{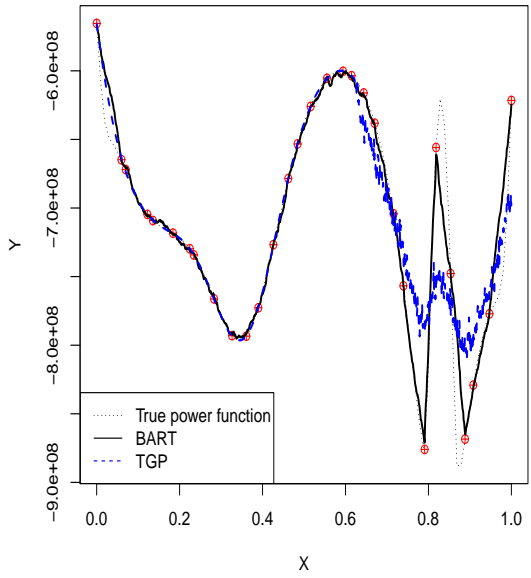}}
\subfigure[Realization~4]{\label{fig:BART-EI-comparison:d} 
\includegraphics[height=3.6in,width=2.95in]{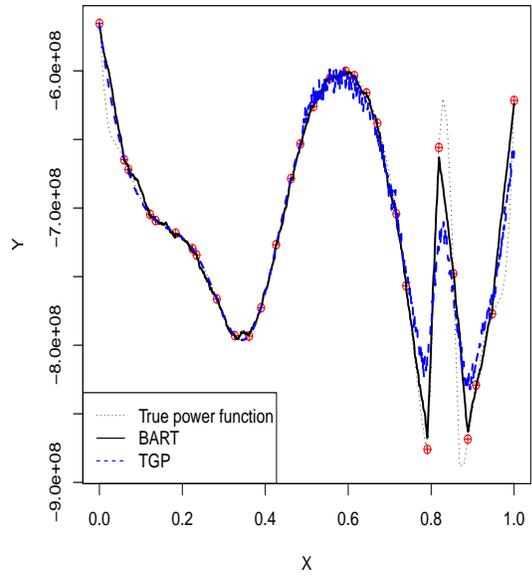}}
\caption{Consistency comparison in fitting surrogate model using BART and TGP for the negative power function in the tidal power example.}
\label{fig:BART-TGP-comparison} 
\end{figure}

A tidal model can provide output of the power (MW) as a function
of the fence location. Thus we have a one-dimensional optimization problem.
The true power function (multiplied by $-1$ so our goal is minimization) is
displayed in each of four panels of Figure~\ref{fig:BART-TGP-comparison}.
Note that in this simplified model the power function can be efficiently
evaluated at many $x$, making the study of emulators more of an academic
exercise. For this particular power function, the effect of a pre-existing
turbine fence at roughly $x=0.83$ is evident, as the power dips (or
negative power spikes) at this location.

The power function is the same in all panels. Each panel displays the
same training set, and the corresponding predictions (posterior
mean at a given $x$) for BART and TGP. The power function displays
considerable nonstationarity (smooth, gradual variation for $0 \leq x \leq
0.65$ and rapid fluctuations for $0.65 \leq x \leq 1$). BART deals more
consistently with this nonstationarity, closely tracking the observed
values. TGP is more likely to declare that this variation is noise, and
the resultant prediction is attenuated, relative to the true function.
Although TGP is able to partition the space into different regions and fit
a GP model in each area, in this case the TGP model is unable to
consistently capture nonstationarity.

\section{Sequential Design with BART and Expected Improvement}\label{sec:bart-ei}

As mentioned in the introduction, we consider a specific kind of sequential
design, in which the goal is to find the optimum of a computer simulator.
In the remainder of the paper, we shall assume that the goal is
minimization, since it is trivial to switch between minimization and
maximization.

We follow the approach of Jones et al. (1998). Let
$f_{\min}=\min\{y_{1},\ldots y_{n_0}\}$
be the current best (i.e.,
smallest) function value among the $n_0$ points sampled thus far.  Jones
et al. (1998) define the improvement at a point $x$ as
$I(x)=\max\{f_{\min}-y(x), 0\}$, which is positive if the (unobserved) response $y(x)$ at location $x$ is less than the current best function value, and $0$ otherwise. They follow the maximum likelihood approach and
take the expectation of this improvement function with respect to $y(x) \sim N(\hat{y}(x), s^2(x))$, where $\hat{y}(x)$ and $s^2(x)$ are the BLUP and the mean square error of response $y$ at $x$. The GP assumption leads to a closed-form expression,
\begin{equation}
\label{eq:JonesEI}
EI(x) = E[I(x)] = (f_{\min}-\hat{y})
 \Phi\left(\frac{f_{\min}-\hat{y}}{s}\right) + s
\phi\left(\frac{f_{\min}-\hat{y}}{s}\right),
\end{equation}
where $\Phi$ and $\phi$ are the standard normal CDF and density.  The first
term (\ref{eq:JonesEI}) captures ``local search'' that seeks to improve the
estimate of the minimum near a currently identified minimum.  The second
term captures ``global search'' which places points in regions where there
is sufficient uncertainty that the minimum response could be nearby.

There are a variety of approaches for evaluating (\ref{eq:JonesEI}) or
similar expressions.  All involve some form of integration over the
predictive distribution for the unobserved response at a new input $x$.
In this paper, we take a fully Bayesian approach, calculating
the expected improvement by taking the
expectation of $I(x)$ with respect to the posterior for all parameters,
using the MCMC samples.  This accounts for both predictive uncertainty and
parameter uncertainty.   We employ a similar approach for TGP and GP.
Although we do not use (\ref{eq:JonesEI}), the ideas of local and global
search are still useful concepts to bear in mind when considering how EI
chooses runs in a sequential design.

The details of the computation of EI are as follows: For BART,
we calculate EI directly from the MCMC output.  For
every posterior draw
$\Theta^{(i)} = (T_1^{(i)}, \ldots, T_m^{(i)}, M_1^{(i)},
\ldots, M_m^{(i)}, \sigma^{(i)})$,
it is straightforward to get the posterior realization of $y(x)$ and hence
$I(x)$. Taking a sample average of $I(x)$ values over the $N$ MCMC draws
yields a MCMC approximation to the Expected Improvement
\begin{eqnarray}
\nonumber EI(x) &=& \frac{1}{N}\sum_{i=1}^N I(x; \Theta^{(i)} )\\
 & =&  \frac{1}{N}\sum_{i=1}^N \max\{f_{\min}-y(x;\Theta^{(i)}),0\},\label{eq:correctEI}
\end{eqnarray}
where $y(x;\Theta^{(i)}) = h(x;\Theta^{(i)})$.  This expression for EI
incorporates uncertainty in all BART parameters.  We use the same MCMC-based
strategy for calculating EI with TGP and GP models.

In order to select a new design point, EI must be maximized over all
possible $x$. The current implementation of BART simultaneously runs
MCMC and generates posterior draws for $h(x)$ at a specified set of $x$
points. Thus, we use space-filling designs (random latin hypercube
designs) to generate a large candidate set, evaluate (\ref{eq:correctEI})
at all points in the candidate set, and select the point with the largest
value of EI as the next input at which to obtain simulator output $y(x)$.
To make comparisons with TGP, we employ the same strategy.

As will be illustrated in the higher dimensional examples in the next
section, this ``candidate set'' strategy for locating the maximizer of EI
becomes computationally challenging as the dimensionality of the input
space grows.

\newcommand{\xtest}{X^{\mbox{\em cand}}}
The sequential design algorithm can thus be summarized as follows:
\begin{enumerate}
\item Obtain an initial design $X_{n_0}$ with $n_0$ points, and evaluate
the simulator at these points, yielding corresponding simulator outputs $Y_{n_0}$.
\item Set iteration $i=0$
\item Select a candidate set $\xtest$ for prediction and evaluation of EI.
\item Fit the model using $X_{n_0+i}, Y_{n_0+i}$.
\item Calculate EI for each point $x \in \xtest$, identifying the maximizer
$x^*$ of EI.
\item Evaluate the simulator at $x^*$, augment $X_{n_0+i}$ and $Y_{n_0+i}$
with $x^*$ and $y(x^*)$, and set $i=i+1$.
\item Unless a stopping condition is met, go to Step 3.
\end{enumerate}
The initial design in Step 1 is a maximin LHD, with $n_0-2$ runs.  The two other
runs are $x=(0,0,...,0)$ and $x=(1,1,...,1)$, which are chosen to aid BART
in assessing uncertainty near the boundaries of the input space.
In Step 3, a new set of candidate points is generated with a random LHD at every iteration of the algorithm.
In Step 4, the ``model'' is BART, TGP or GP.
In Step 5 EI is calculated using MCMC integration.
The ``stopping condition'' in Step 7 could either be a convergence
criterion or tied to a budget.  In the remainder of the
article, we shall assume a fixed budget of a total of $n =
n_{new} + n_0$ evaluations of the simulator.


\section{Experimental Results}\label{sec:results}

In this section, we demonstrate the performance of our proposed approach
using a few simulated examples and the tidal power application discussed in
Section~\ref{sec:tidal_power}. First a simple one-dimensional example is
used to illustrate the sequential design approach with BART as the
emulator. Then, we demonstrate the proposed approach on more complex
simulators with one or more inputs. In all the examples presented here, our
objective is to find the global minimum.

The average performance of BART in the sequential design framework is
compared with TGP, a popular tree based surrogate for emulating
non-stationary computer simulators.  We also make compraisons with a
standard GP model.  The performance of the sequential techniques is also compared with non-sequential ``one-shot" space-filling designs. Under the one-shot design setup, the simulator outputs based on an $(n_0 + k)$-point random maximin LHD was used to
estimate the global minimum, where $0\le k \le n_{new}$. Although
a series of different LHDs of increasing size is not a viable sequential
design strategy, it does provide a basis for comparison. All results
presented in this section are based on 100 realizations (i.e., 100
different starting designs based on random maximin LHDs, each followed by
sequential design generation). For convenience, the EI-based sequential design approach using
BART with (\ref{eq:correctEI}) is referred to as BART-EI, and TGP-EI
(GP-EI) refers to the sequential procedure if TGP (GP, respectively)
is used instead of BART.

Implementations of BART-EI, TGP-EI and GP-EI follow the algorithm outlined at
the end of Section 3. For all examples presented here, BART runs follow the
prior parameters and operational settings described in
Sections~\ref{sec:bartprior}, whereas TGP parameter settings are
discussed in \ref{sec:tgp}.  The GP settings are the same as TGP, but with
the constraint that there be a single node, giving the same GP over the
entire input space.  Both TGP and GP also use MCMC integration to
evaluate EI.

For all examples, the following details apply.  We start with fitting a
surrogate to the simulator data on a $n_0$-point random maximin Latin
hypercube design (LHD). These designs are generated using the R function
{\tt maximinLHS} (Carnell, 2009)
which takes random starting points and hence the
output designs are random. Jones et al. (1998) and Loeppky et al. (2009) suggest using approximately $n_0=10d$ points as a reasonable rule-of-thumb for an initial design. However, the optimal choice of $n_0$ varies with the complexity of the computer simulator. The remaining $n_{new} = n-n_0$ points are sequentially added one at-a-time by optimizing the EI criterion and refitting the surrogate (BART, TGP or GP). The size of the candidate sets $\xtest$ for evaluation of EI at unsampled input locations varies with the complexity of the simulators.

We use the running best estimate of the global minimum (i.e., the smallest
$y$ value observed thus far) as our performance measure. An alternative
considered, but not reported here, is the distance between the input locations  of the true and estimated global minimum.  This may not be useful if the simulator has multiple global minima, which is the case in Example~3.\\

\begin{figure}[h!]\centering
\subfigure[BART fit with the initial design]{\label{fig:eg1-BART-EI:a} 
\includegraphics[width=3.25in]{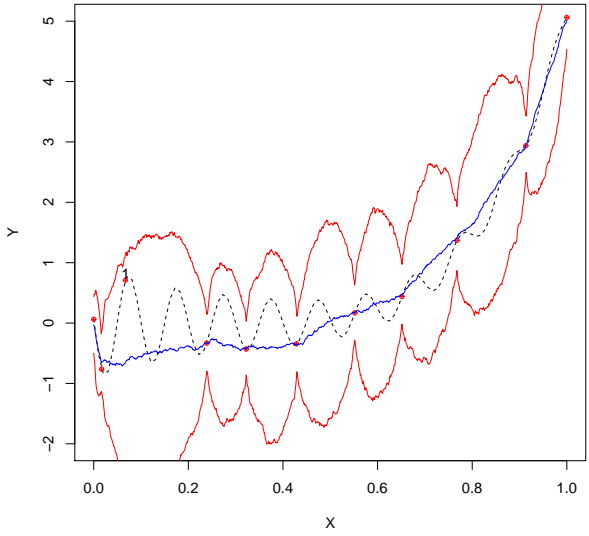}}\hspace{-0.5cm}
\subfigure[BART fit after adding 5 design points]{\label{fig:eg1-BART-EI:b} 
\includegraphics[width=3.25in]{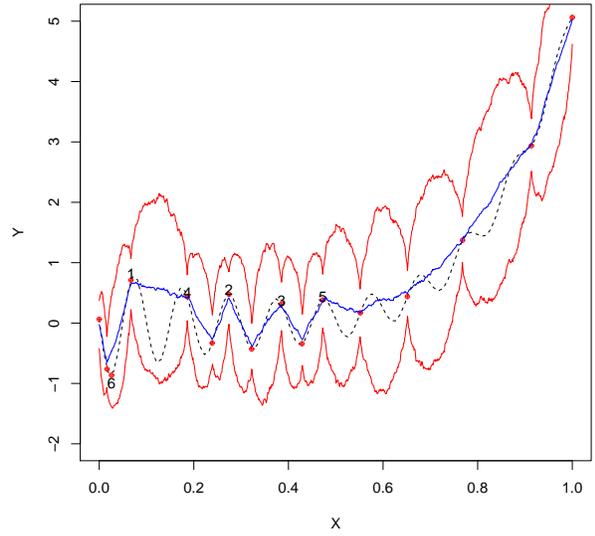}}
\subfigure[BART fit after adding 10 design points]{\label{fig:eg1-BART-EI:c} 
\includegraphics[width=3.25in]{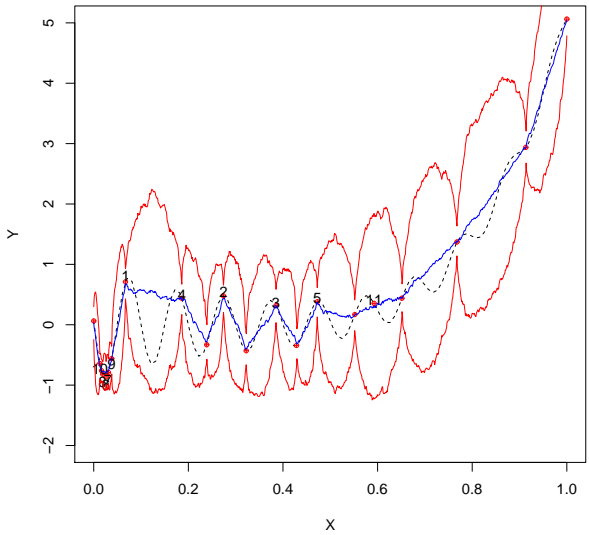}}\hspace{-0.5cm}
\subfigure[BART fit after adding 15 design points]{\label{fig:eg1-BART-EI:d} 
\includegraphics[width=3.25in]{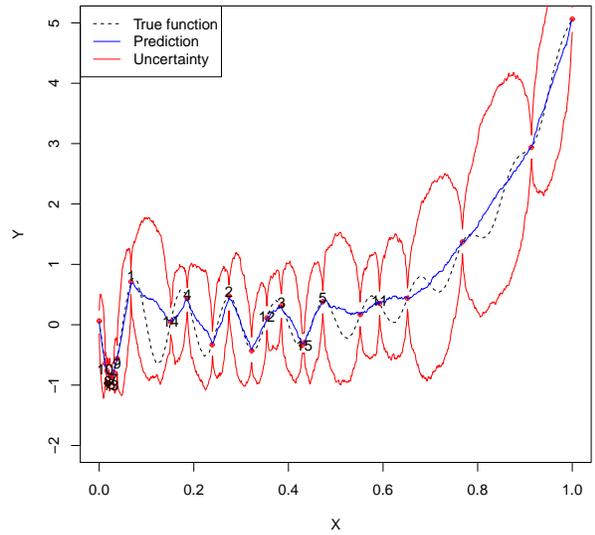}}
\caption{The BART-EI procedure for the one dimensional simulator
(\ref{eq:testfn_1d}). The uncertainty bounds are $\hat{y}(x) \pm 2s(x)$ for
$0\le x\le 1$, and the numbers indicate the order in which follow-up design
points are placed.}
\label{fig:BART-EI} 
\end{figure}

\textbf{Example 1.} Suppose the simulator outputs are generated using a
the simple one-dimensional test function
\begin{equation}
y(x) = \frac{\sin(10 \pi x)}{ 2 x} + (x-1)^4, \quad x \in [0.5, 2.5].
\label{eq:testfn_1d}
\end{equation}
The inputs are scaled to $[0, 1]$ for implementation. Gramacy and Lee
(2012) used this example to demonstrate the benefits of including a nugget
in a GP model. Figure~\ref{fig:BART-EI} illustrates the sequential
BART-EI methodology for $n_0=10$ and $n_{new} = 15$ runs.

Although BART does not accurately capture all high-frequency oscillations
of the function, it appears to effectively guide sequential
design for the minimum, utilizing a combination of local and global search.
Figure~\ref{fig:eg1-BART-EI:a} shows that the first follow-up trial is in
the vicinity of the current best estimate of the global minimum (i.e.,
supporting the local search). Figure~\ref{fig:eg1-BART-EI:b} shows that the
next four follow-up points support exploration and the points are away from
the running global minimum.  Most additional points are near the running global minimum (see panel (c)), although some points such as 11, 12, 14 and 15 are placed in regions of high uncertainty (i.e., global search). The emulator constructed using BART exhibits ``football shaped" uncertainty bounds as typically shown by GP
models for deterministic simulators.  The variance of the predicted
response at the training points is not zero, however, due to the inclusion
of a noise term in the BART model.

\begin{figure}[h!]\centering
\includegraphics[height=4in,width=5.5in]{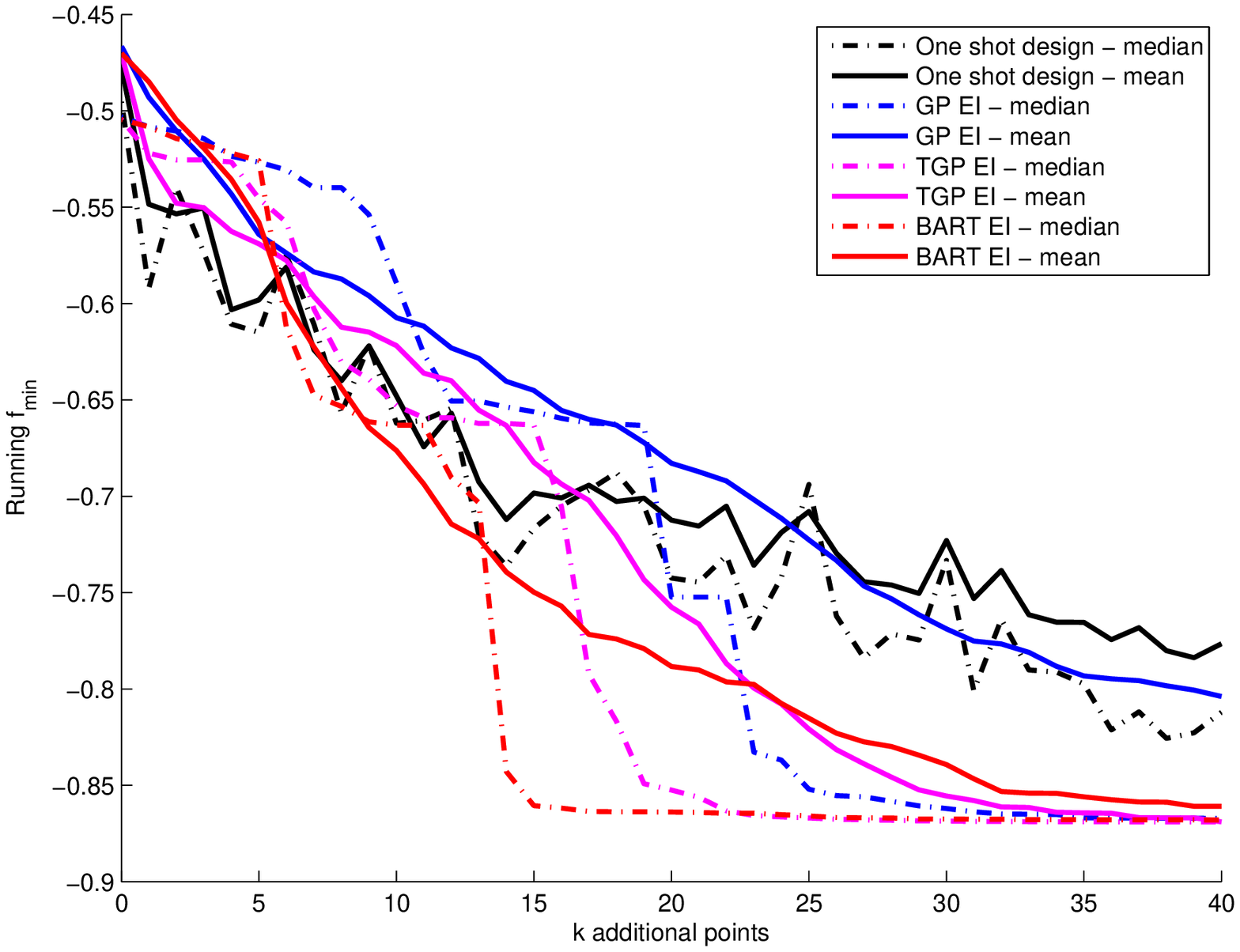}
\caption{Comparison of the running best estimate of the global minimum
using BART-EI, TGP-EI and one-shot approach for the one dimensional
simulator (\ref{eq:testfn_1d}), with $n_0=10$ and $n_{new}=40$. }\label{fig:eg1_ymin}
\end{figure}

To compare GP, TGP and BART, we repeated the experiment 100 times.
Following the $n_0=10d$ rule of thumb, we started with $n_0=10$ points and added $n_{new}=40$ runs sequentially one at-a-time. We used 1000-point random
LHDs as candidate sets ($\xtest$) at which EI was evaluated to choose the next run.
Figure~\ref{fig:eg1_ymin} presents the performance comparison
between BART-EI, TGP-EI, GP-EI and the one-shot design method.
Performance is summarized using mean and median
(over 100 realizations) of the running best estimates of the global
minimum for each method.

As expected, the sequential design approaches outperform the one-shot
design scheme. BART-EI slightly outperforms TGP-EI and is clearly superior
to GP-EI. The mean value of the running best estimate is typically larger than the
median because the best estimate is bounded below by the global minimum of
y(x).  Thus, the horzontal lines just below -0.85 indicate that the global
minimum has been found in over half of the 100 replicates.  The more
gradual approach of the mean lines indicates that in some cases, the global
optimum has not yet been found.\\

\begin{figure}[h!]\centering
\includegraphics[height=4.5in,width=5.5in]{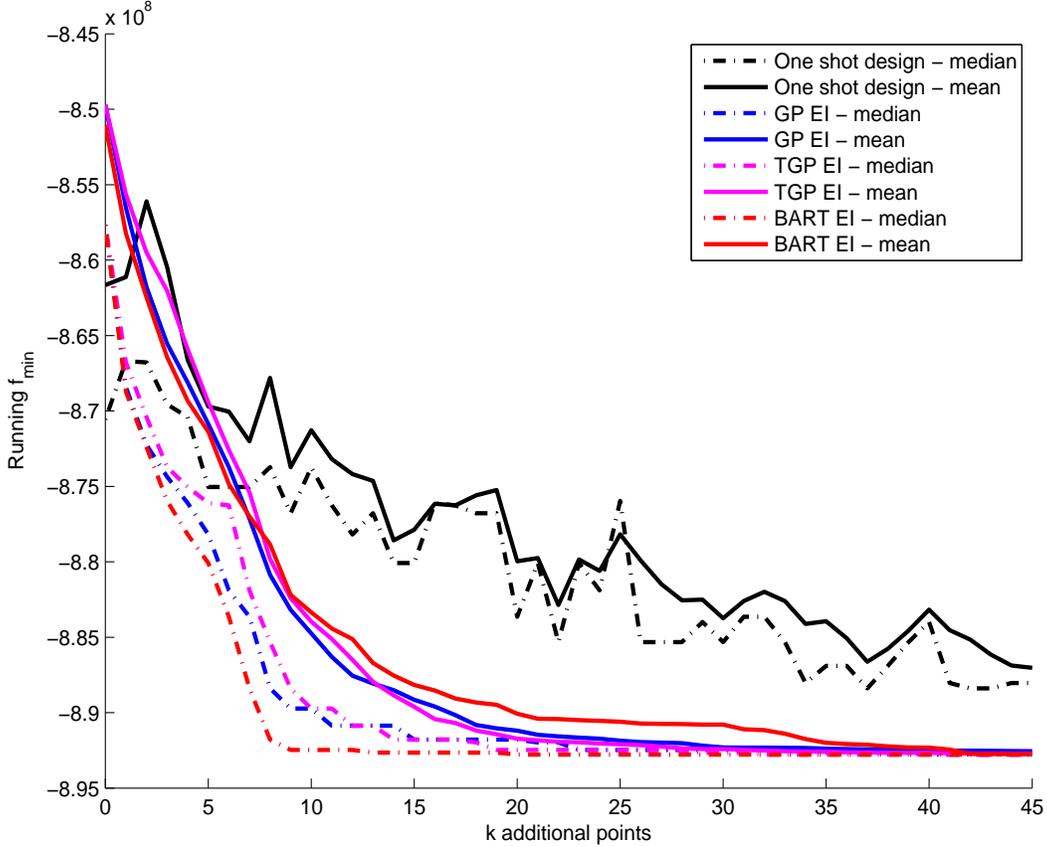}
\caption{Comparison of the running best estimate of the global minimum
using BART-EI, TGP-EI and one-shot approach for the tidal power simulator,
with $n_0=15$ and $n_{new}=45$.}\label{fig:eg2_tidal_ymin}
\end{figure}

\textbf{Example 2.} We revisit the tidal power example, where our objective
is to maximize the power function (or equivalently, minimize the negative
power function). Recall from Figure~\ref{fig:BART-TGP-comparison} that the
negative power function is very spiky near the global minimum and there is
another spike with the function value close to the global minimum. This
makes the optimization problem tricky especially if the design points are
not densely sampled in both the spikes. Thus, we started the sequential
algorithms with a slightly larger initial design, $n_0=15$, and added $n_{new}=45$ follow-up points one at-a-time. The EI criterion was calculated on a 1000-point random LHD.  Figure~\ref{fig:eg2_tidal_ymin} summarizes the mean and median (over 100 realizations) of the running best estimates of the global minimum.

As in Example~1, the sequential design approaches outperform the
naive one-shot approach. But, the performance comparison of BART-EI
and TGP-EI shows interesting results. The mean curve for BART-EI
is converging at a slower rate than TGP-EI or GP-EI, but in terms of median
performance, BART dominates. It
turns out that BART-EI can sometimes get stuck in the wrong spike
if there are not enough design points in the regions of interest (see
Figure~\ref{fig:BART-TGP-comparison}).
 \\

\begin{figure}[h!]\centering
\includegraphics[height=4.5in,width=5in]{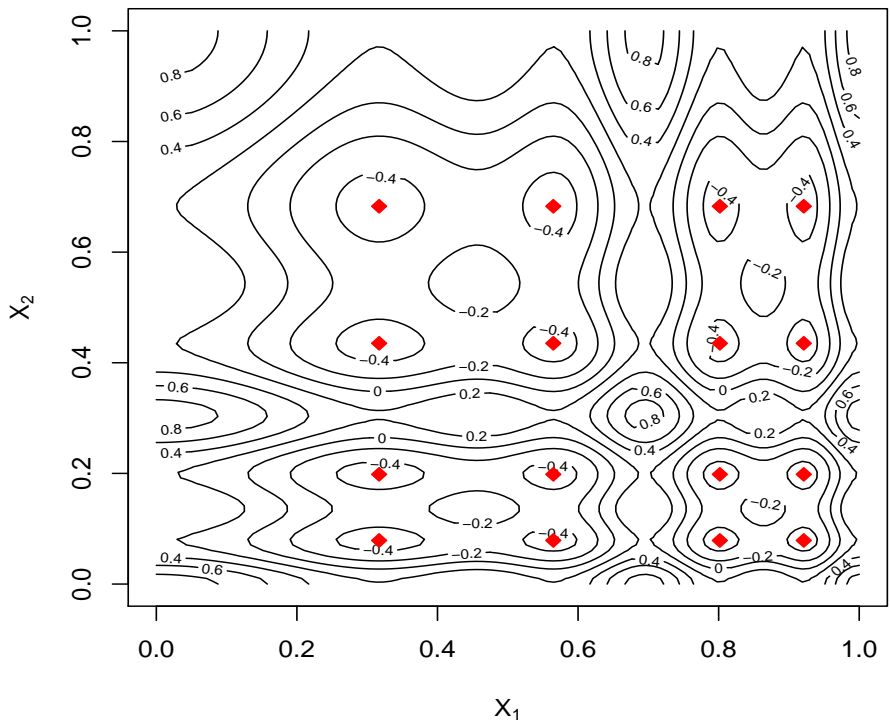} \vspace{-0.75cm}
\caption{Contour plot of the two-dimensional test function shown in (\ref{eq:testfn_2d}).}\label{fig:eg3_contour}
\end{figure}

\textbf{Example 3.} Let $x=(x_1, x_2) \in [0, 1]^2$, and the simulator outputs be generated from a two-dimensional additive function given by (R\"{o}nkk\"{o}nen et al. 2008)
\begin{equation}
y(x) = \frac{1}{4}\sum_{i=1}^2 [\cos(4\pi w_i) + \alpha \cdot \cos(8\pi w_i)],
\label{eq:testfn_2d}
\end{equation}
where  $\alpha = 0.8$ and $w_i = \sum_{j=0}^{n_i} {n_i \choose j}P_{i,j}(1-x_i)^{n_i-j}x_i^j$ for $i=1,...,d$, $n_1=n_2=4$ and $P_1 = (0,0.1,0.2,0.5,1), P_2 = (0,0.5,0.8,0.9,1)$. The contour plot in Figure~(\ref{fig:eg3_contour}) shows that the test function has 16 global minima with $y_{min} \approx -0.478$ (shown in red diamonds).

\begin{figure}[h!]\centering
\includegraphics[height=4.5in,width=5.5in]{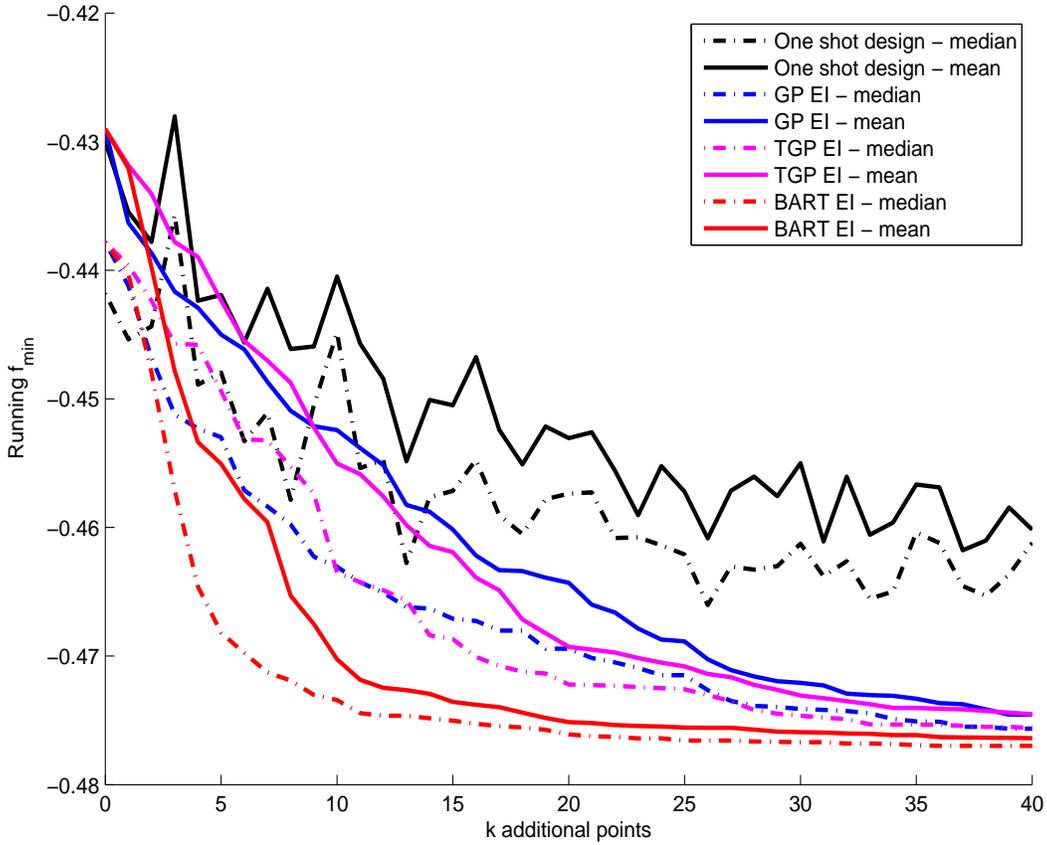}
\caption{Comparison of the running best estimate of the global minimum
using BART-EI, TGP-EI and one-shot approach for the two dimensional
simulator (\ref{eq:testfn_2d}), with $n_0=20$ and $n_{new}=40$.}\label{fig:eg3_ymin}
\end{figure}

Since the function has multiple global minima, one might suspect that
the estimation of the global minimum would be relatively easier. But
the local versus global search trade-off feature of the EI criterion
forces the follow-up trials to jump from the neighbourhood of one global
minimum to another. That is, instead of precisely estimating one global
minimum, the EI criterion tries to minimize the prediction uncertainty
near several global minima. Consequently, the sequential algorithms
require a large number of follow-up trials to accurately estimate the
global minimum. Following the $n_0=10d$ rule, we started the implementation with $n_0=20$ points chosen using random maximin LHD and sequentially added $n_{new}=40$ follow-up
points using the EI criterion which was evaluated at 5000-point random LHD
over $[0, 1]^2$. Figure~\ref{fig:eg3_ymin} presents the mean and median
(over 100 simulations) of the running best estimate of the global minimum.

It is clear from Figure~\ref{fig:eg3_ymin} that BART-EI performs
significantly better than TGP-EI and GP-EI.
The fact that the spikes are aligned to the axes should help both BART and
TGP.  TGP should benefit more from the smoothness of the response function,
while BART may be better suited to identify the additive structure.
The slower convergence of TGP-EI can perhaps be
attributed to inconsistent prediction of TGP.

In this example a larger candidate set $\xtest$
was used to evaluate EI and select the next run.  Recall from
Section~\ref{sec:bart-ei} that the EI criterion is evaluated only at
these cadidate points, and thus their distribution and density play a role
in estimating the global minimum.  With a two dimensional input space, we
elected to use more candidate points.\\

\begin{figure}[h!]\centering
\includegraphics[height=4in,width=4.5in]{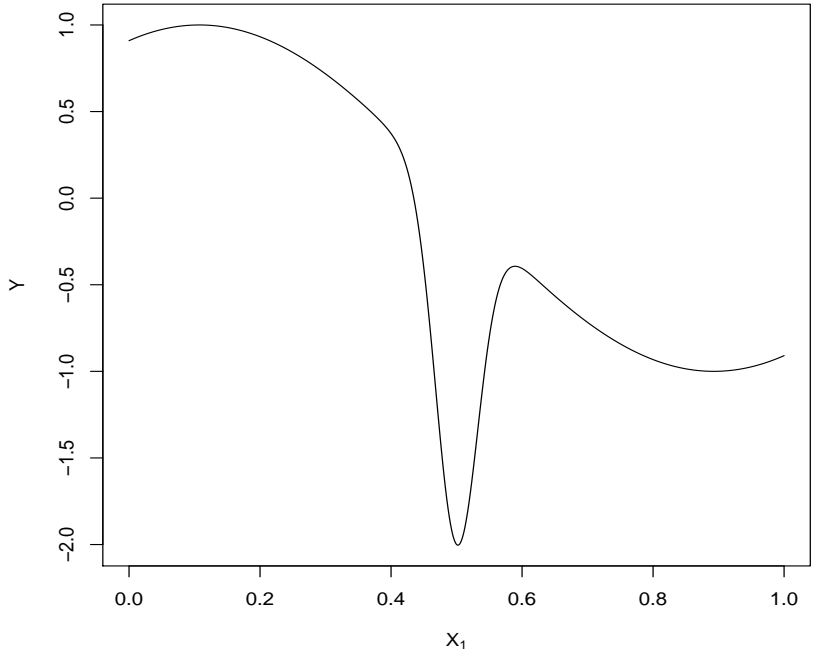}
\caption{True one-dimensional test function used to build the simulator in (\ref{eq:testfn_4d}).}\label{fig:eg4_1d}
\end{figure}

\textbf{Example 4.} Suppose the computer simulator outputs are generated using the four-dimensional test function given by
\begin{equation}
y(x) = \sum_{i=1}^4 -\sin(x_{i}) - 2 \exp\left(-30x_{i}^2\right), \quad x_i \in [-2,2],
\label{eq:testfn_4d}
\end{equation}
where the input variables, $x = (x_1, x_2, x_3, x_4)$, are scaled to $[0,
1]^4$. This test function is based on a one-dimensional function from
DiMatteo, Genovese and Kass (2001), and has a unique global minimum with
$y_{min} = -8$, but the global minimum is in a narrow spike.
Figure~\ref{fig:eg4_1d} shows the one-dimensional function. The detection of this spike in the four dimensional $[0, 1]^4$ space would require at least a few design points in $[0.4, 0.6]^4$, that is in $0.16\%$ of the total volume, otherwise the surrogate models can get misled by the overall shape (i.e, excluding the spike).

\begin{figure}[h!]\centering
\includegraphics[height=4.5in,width=5.5in]{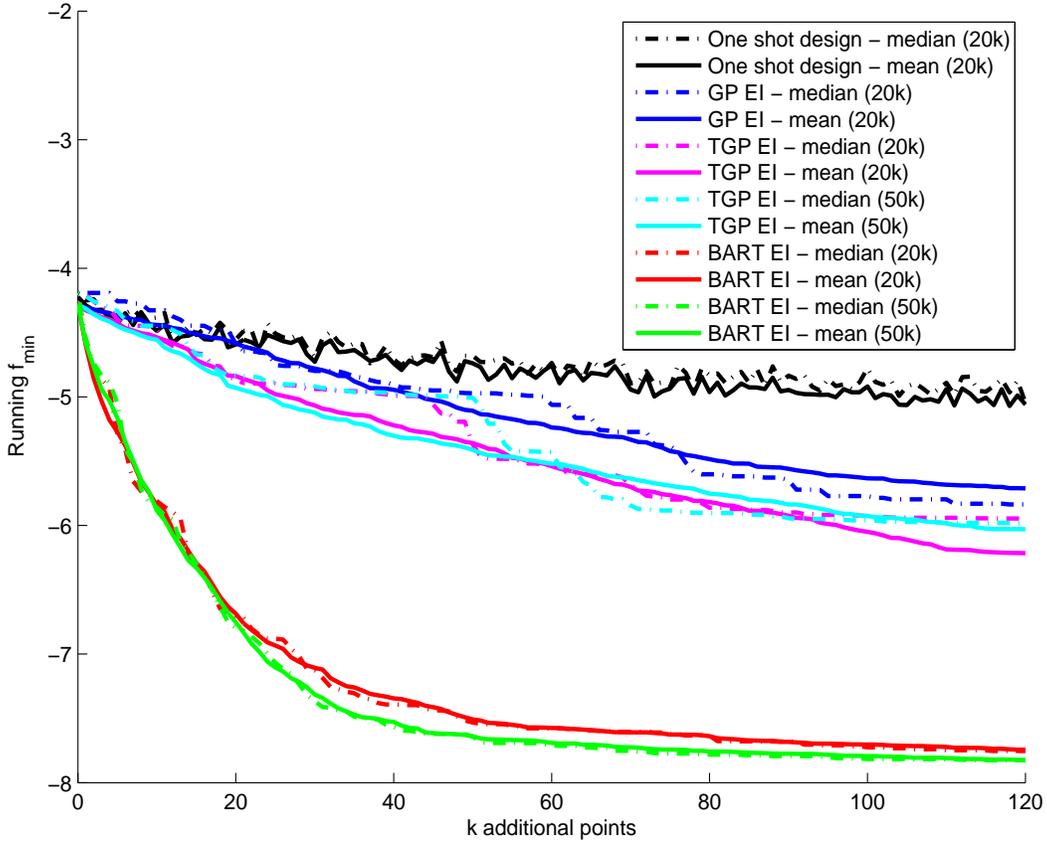}
\caption{Comparison of the running best estimate of the global minimum
using BART-EI and TGP-EI for the four dimensional simulator
(\ref{eq:testfn_4d}), with $n_0=30$ and $n_{new}=120$.}\label{fig:eg4_ymin}
\end{figure}

A space-filling LHD-based initial design would have to be very large to guarantee a few points in the spiky region $[0.4, 0.6]^4$. On the other hand, if we ignore this spike, the test function or simulator is relatively simple to emulate. We start the sequential procedure with only $n_0=30$ points (smaller than the recommended $n_0=10d$ rule) and leave the discovery of the spiky region to the follow-up runs. We sequentially added $n_{new}=120$ follow-up points. To ensure the evaluation of the EI criterion in the spiky region, we considered dense candidate sets $\xtest$ with 20,000 and 50,000 points. Figure~\ref{fig:eg4_ymin} summarizes the running global minimum.

BART-EI very clearly outperforms the other methods, with the median curves
indicating that the global optimum is attained or nearly attained with an
additional 50 points, in over 50\% of the 100 replicates. Evidently, BART
is much better suited to exploit the additive structure in
(\ref{eq:testfn_4d}). There is a small but non-negligible benefit to evaluating EI criterion for BART with a larger (50K) candidate set.

\section{Discussion}
The various examples in Section 4 demonstrate that BART can be an effective
engine for sequential design and optimization.  In situations
where a GP model or perhaps localized GP models are appropriate surrogates,
optimization with BART is still competitive.  When there
are nonstationarities, abrupt changes or additive structure, we see that
BART can deliver the optimum with fewer runs of the simulator.

As in any sequential design procedure, the size and configuration
of points in the initial design can impact the performance of the
sequential design. Jones et al. (1998) and Loeppky et
al. (2009) suggest using approximately $n_0=10d$ points as a reasonable
rule-of-thumb for an initial design. However, the optimal choice of $n_0$
depends on the complexity of the simulator. Certainly if a very small
number of runs are used in the initial design, the algorithm based on
BART (or TGP) may take longer to find the optimum. On the other hand,
one may not want use a very large initial design and put several points
in the unimportant regions of the input space.

The reader may notice that the BART model is not continuous, but rather
piecewise constant.  However, by utilizing many trees in the sum ($m=100$ in our
examples) and also employing posterior averaging over these trees, the BART
model builds a surrogate that can have many small steps, providing an
effective approximation to a continuous function.  In our experience, the
BART model seems effective at deciding where to probe the simulator (i.e.
carry out additional runs), even in situations where it may not provide as
``nice'' a fit as a GP model.  Of course, if a smooth prediction is desired (and
possibly a better extrapolation to the minimum, if there is smoothness
nearby) one may train a GP or TGP model on the set of all simulator
runs, once the sequential design procedure has terminated.  If, on the
other hand, there is a possibility of nonstationarity, abrurpt changes or
even discontinuities in the response function, then BART seems more likely
to be an effective engine for sequential design and optimization.

As observed in Section 3, to evaluate EI (\ref{eq:correctEI}), we must
specify (or design) a candidate set of points at which to generate predictions
from the BART model.  Since this set can be large (e.g., 20,000 to 50,000
in the 4-dimensional example), we used a random LHD.  A maximin LHD (or other space-filling design) with thousands of points could be computationally challenging to identify.

Also related to the evaluation of EI, for the sake of comparison we used a
``batch predict'' approach to both BART and TGP.  The current
implementation of BART is not equipped with a ``predict'' function than
allows predictions for new inputs.  Instead the predictions are generated
during the MCMC estimation of the model.  To make comparisons with TGP, we
thus used the same ``batch predict'' strategy.  We are currently exploring
alternate ways to evaluate EI and find the most promising points to add
during the sequential design.\\

\begin{center} \textbf{ACKNOWLEDGMENTS}
\end{center}
We thank Richard Karsten for providing the tidal power model. The work
of Wang was supported by Acadia Honours Student Research Award, and the
work of Chipman and Ranjan were supported by Discovery grants from the
Natural Sciences and Engineering Research Council of Canada.  Computing
resources were provided by the Acadia Centre for Mathematical Modelling and
Computation.\\


\begin{center}
{\textbf{REFERENCES}}
\end{center}

\begin{description}
\item {\sc Carnell, R.} (2009), ``lhs: Latin Hypercube Samples'', R package version 0.5.

\item {\sc Chipman, H.A., George, E.I., \normalfont{and} \sc McCulloch, R.E.} (2010),   ``BART: Bayesian Additive Regression Trees", \emph{Annals of Applied Statistics}, 4, 266--298.

\item {\sc DiMatteo, I. Genovese, C.R. \normalfont{and} \sc Kass, R.E.} (2001), ``Bayesian curve fitting with free-knot splines", \emph{Biometrika}, 88:1055--1071.

\item {\sc Gramacy, R.B. \normalfont{and} \sc Lee, H.K.H.} (2008). ``Bayesian Treed Gaussian Process Models with an Application to Computer Modeling", \emph{Journal of the American Statistical Association}, 103, 1119--1130.

\item {\sc Gramacy, R.B. \normalfont{and} \sc Lee, H.K.H.} (2012). ``Cases for the Nugget in Modelling Computer Experiments'', \emph{Statistics and Computing}, 22, 713--722.

\item {\sc Jones, D., Schonlau, M., \normalfont{and} \sc Welch, W.} (1998).  ``Efficient Global Optimization of Expensive Black-Box Functions".  \emph{Journal of Global Optimization}, 13, 455 - 492.

\item {\sc Karsten, R., McMillan, J., Lickley, M. \normalfont{and} \sc Haynes, R.} (2008). ``Assessment of tidal current energy for the Minas Passage, Bay of Fundy". \emph{Proceedings of the Institution of Mechanical Engineers, Part A: Journal of Power and Energy}, 222, 493 - 507.

\item {\sc Loeppky, J. L., Sacks, J. \normalfont{and} \sc Welch, W. J.} (2009). ``Choosing the Sample Size of a Computer Experiment: A Practical Guide", \emph{Technometrics}, 51(4), 366--376.

\item {\sc Ranjan, P., Haynes, R. \normalfont{and} \sc Karsten, R.} (2011), ``A Computationally Stable Approach to Gaussian Process Interpolation of Deterministic Computer Simulation Data", \emph{Technometrics}, 53(4), 366 -- 378.

\item{\sc Rasmussen, C. E. \normalfont{and} \sc Williams, C. K. I.} (2006). \emph{Gaussian Processes for Machine Learning}. The MIT Press.

\item {\sc R\"{o}nkk\"{o}nen, J., Li, X., Kyrki, V. \normalfont{and}\sc Lampinen, J. A.} (2008), ``Generator for Multimodal Test Functions with Multiple Global Optima". Proceedings of 7th International Conference on Simulated Evolution and Learning (SEAL'08), Melbourne, Australia, 239--248.

\item {\sc Santner, T. J., Williams, B. J., \normalfont{and} \sc Notz, W. I.} (2003), ``The Design and Analysis of Computer Experiments," \emph{Springer-Verlag, New York.}

\item {\sc Wang, W.} (2010), ``Modelling Energy Output to Optimize Tidal Turbine Placement", B.Sc. Thesis, Acadia University, Canada.
\end{description}

\end{document}